\documentclass[aps,prd,preprintnumbers,nofootinbib]{revtex4}

\usepackage[dvipdfmx]{graphicx}
\usepackage{bm,latexsym,amsmath,amsthm,amssymb,amsfonts,color,mathrsfs}

\usepackage[normalem]{ulem}

\newcommand{\tr}{\tilde{r}}

\begin{document}

%<<<<<<<<<<<<< TITLE >>>>>>>>>>>>>>>%
\title{\uline{}Exact solutions for static spherically symmetric spacetime with a perfect fluid in Rastall theory}

%<<<<<<<<<<<<< AUTHOR  >>>>>>>>>>>>>>>%
\author{${}^{1}$Yoshimune Tomikawa, ${}^{2}$Ichika Obara, and ${}^{2}$Takumi Orimo}

%<<<<<<<<<<<<< AFFILIATION >>>>>>>>>>>>>>>%
\affiliation{${}^{1}$Division of Science, School of Science and Engineering, Tokyo Denki University, Saitama 350-0394, Japan}
\affiliation{${}^{2}$Graduate School of Science and Engineering, Tokyo Denki University, Saitama 350-0394, Japan}

%
%======================================%
%<<<<<<<<<<<<< ABSTRACT >>>>>>>>>>>>>>>%
%======================================%
%
\begin{abstract}
In general relativity, exact Liouvillian solutions for a static and spherically symmetric spacetime 
with a perfect fluid and the equation of state $p(r)=w\rho (r)$
are known only for $w=0,-\frac{1}{6}, -\frac{1}{5}, -\frac{1}{3}$, and $-1$.
We extend this setup to Rastall theory, presenting the relation between the Rastall parameter and the constant $w$,
and deriving exact solutions that correspond to their known counterparts in general relativity, except for $w=-\frac{1}{6}$.
Furthermore, we find that, when $w\neq \frac{1}{3}, -1$, 
there exist several types of solutions whose behaviors change depending on the choice of constants.
\end{abstract}

\maketitle

%
%======================================%
%<<<<<<<<<<<< SECTION I  >>>>>>>>>>>>>>%
%======================================%
%
\section{Introduction} 

Rastall gravity theory is a modified gravity theory proposed by Rastall in 1972 \cite{Rastall1972}.
It is a theory in which the covariant divergence of the stress-energy tensor depends on the curvature,
and it is characterized by the violation of the conservation law for the stress-energy tensor.
Although this theory is of interest and has been the subject of many studies
\cite{LH1982, Smalley1983, Visser2018, DMLHC2018, CLO2023, Golovnev2024, RT1996, BDFPR2012, 
OVFC2015, KD2019, MJAAA2024, AS2018, SWJEA2023, WN2024, WANJA2025,
Smalley1984, SN2017, MS2019, SH2020, FPR2023, PFZ2026, HMD2017, 
HD2017, OVF2016, BFPS2016, HBC2018, Waseem2024, HBC2019},
there remain unresolved problems and differing opinions regarding it. 

One of the problems is whether Rastall theory is equivalent to general relativity.
This equivalence problem was pointed out by Lindblom and Hiscock in 1982 \cite{LH1982}, 
and was later revisited by Visser in 2018 \cite{Visser2018}.
Visser's argument for this equivalence is that the field equations of Rastall theory
can be transformed into a form equivalent to general relativity 
by redefining the stress-energy tensor \cite{Visser2018}.
In opposition to this, Darabi et al. \cite{DMLHC2018} and Chagoya et al. \cite{CLO2023} have argued
that while Visser's argument is correct as a mathematical transformation of the equations, 
physical problems still remain.
Conversely, Golovnev supported Visser's argument \cite{Golovnev2024}.
Another problem is that the action for Rastall theory remains unknown.
Although attempts have been made to formulate it and several candidates for the action have been proposed 
\cite{Golovnev2024, Smalley1984, SN2017, MS2019, SH2020, FPR2023, PFZ2026}, 
one fails to yield the field equations \cite{SN2017}, 
while the other is considered to have an unnatural construction \cite{Golovnev2024}.

%%%%%%%%%%%%%%%%%%%%%%%%%%%%%%%%%%%%%%%%%%%%%%%%%%%%%%%%%%%%%%%%%%%%%%%%%%%%%%%%

On the other hand, from the cosmological perspective, Rastall theory yields results of interest 
\cite{RT1996, BDFPR2012, OVFC2015, KD2019, MJAAA2024, AS2018, SWJEA2023, WN2024, WANJA2025}.
For instance, Rastall cosmology includes the $\Lambda$CDM model \cite{BDFPR2012, OVFC2015, KD2019, MJAAA2024}
or various other models \cite{AS2018, SWJEA2023, WN2024, WANJA2025}.
Motivated by these findings, numerous studies have also explored solutions 
to the field equations in Rastall theory \cite{OVF2016, BFPS2016, HMD2017, HD2017, HBC2018, Waseem2024, HBC2019}.
For instance, Ref. \cite{BFPS2016} obtained exact solutions for a static and spherically symmetric spacetime with a scalar field.
References \cite{HMD2017, HD2017} derived static and spherically symmetric black hole solutions with a perfect fluid
under an additional assumption on the metric (namely, $g_{tt}g_{rr}=-1$ and $g_{AB}=r^2 \sigma_{AB}$, 
where $A,B$ are angular components and $\sigma$ is the metric of the 2-dimensional unit sphere).
Reference \cite{HBC2018} derived the Tolman metrics in Rastall theory and analyzed their 
behaviour (see also Ref. \cite{Waseem2024}).
Furthermore, wormhole solutions were investigated in Ref. \cite{HBC2019}.

Studies on the derivation of exact solutions are also relevant to the equivalence problem. 
That is because if a solution is derived that is fundamentally different from those in general relativity,
it would serve as a counterexample.
In this paper, we adopt the equation of state $p(r)=w\rho (r)$ and 
derive exact solutions for a static and spherically symmetric spacetime with a perfect fluid.
For this purpose, we find specific relations between the Rastall parameter and the constant $w$ to solve the field equations.
In addition, the derived solutions correspond to their counterpart solutions in general relativity
for the values $w=0,-\frac{1}{5}, -\frac{1}{3}$, and $-1$.
We can see that this enables us to establish a correspondence with the Rastall parameter for these values of $w$.
In general relativity, there exist exact Liouvillian solutions for a static and spherically symmetric (4-dimensional) spacetime 
with a perfect fluid and the equation of state $p(r)=w\rho (r)$ 
if and only if $w=0,-\frac{1}{6},-\frac{1}{5}, -\frac{1}{3}$, or $-1$ \cite{Ivanov2002, IL2014}.
Here, roughly speaking, {\it Liouvillian} means that functions can be expressed 
in terms of elementary functions and a finite number of their integrals.
In particular, if a differential equation can be solved by quadratures, its solution is Liouvillian.
In this paper, we derive exact solutions for all these values except for $w=-\frac{1}{6}$,
and show that these exact solutions are the same as those in general relativity 
if the Rastall parameter for each solution is chosen appropriately.
Specifically, we show that the solution reduces to the Minkowski spacetime when $w=0$,
and the Schwarzschild-(anti) de Sitter spacetime when $w=-1$.
Furthermore, we obtain a solution of the same form as that in Ref. \cite{Semiz2022} when $w=-\frac{1}{5}$,
and two types of solutions of the same forms as those in Ref. \cite{CSS2002} when $w=-\frac{1}{3}$.

%%%%%%%%%%%%%%%%%%%%%%%%%%%%%%%%%%%%%%%%%%%%%%%%%%%%%%%%%%%%%%%%%%%%%%%%%%%%%%%%

The rest of this paper is organized as follows. 
In Sec. \ref{sec-setup}, we describe the setup for a static and spherically symmetric spacetime 
with a perfect fluid and the equation of state $p(r)=w\rho (r)$.
In Sec. \ref{sec-sol}, we derive exact solutions that correspond to 
their counterparts in general relativity for the values $w=-1, 0,-\frac{1}{5}$ 
by specifying the relation between the Rastall parameter and the constant $w$.
Similarly, in Sec. \ref{sec-exact}, we obtain the corresponding exact solutions for $w=-\frac{1}{3}$.
In addition, we find that when $w\neq \frac{1}{3}, -1$, 
there exist several types of solutions whose behaviors change depending on the choice of constants.
Finally, we provide a summary and discussion
in Sec. \ref{sec-summary}, and present additional calculations in Appendix \ref{sec-app}.
Throughout the paper, we use units in which the speed of light and the Newtonian constant of gravitation are 
set to unity, $c=1$ and $G=1$. The metric sign convention is $(-,+,+,+)$.

%======================================%
%<<<<<<<<<<<< SECTION II  >>>>>>>>>>>>>%
%======================================%
\section{Setup} \label{sec-setup}

In Rastall theory, the gravitational field equation is given by
\begin{eqnarray}
R_{\mu \nu}+\left( \kappa \lambda -\dfrac{1}{2} \right) g_{\mu \nu} R =\kappa T_{\mu \nu}, \label{Eineq-Rastall}
\end{eqnarray}
and the non-conservation law for the stress-energy tensor,
\begin{eqnarray}
\nabla_{\mu} T^{\mu}_{\nu} =\lambda \nabla_{\nu} R, \label{energy-Rastall}
\end{eqnarray}
is assumed \cite{Rastall1972}.
Here, $\lambda$ is the Rastall parameter, which is a free parameter, and $\kappa$ is a constant satisfying
$(6\kappa \lambda -1) \kappa =8\pi (4\kappa \lambda -1)$.

The contraction of Eq. (\ref{Eineq-Rastall}) yields
\begin{eqnarray}
(4\kappa \lambda -1) R=\kappa T.
\end{eqnarray}
Therefore, if $\lambda \neq \frac{1}{4\kappa}$, 
the gravitational field equation (\ref{Eineq-Rastall}) can be rewritten as
\begin{eqnarray}
R_{\mu \nu}-\dfrac{1}{2} g_{\mu \nu} R =\kappa \left( T_{\mu \nu} -\dfrac{a-1}{2} g_{\mu \nu} T \right), \label{Eineq-a}
\end{eqnarray}
and the non-conservation law (\ref{energy-Rastall}) becomes
\begin{eqnarray}
\nabla_{\mu} T^{\mu}_{\nu} =\dfrac{a-1}{2} \nabla_{\nu} T, \label{energy-a}
\end{eqnarray}
where $a:=\frac{6\kappa \lambda -1}{4\kappa \lambda -1}$ \cite{BFPS2016}.
Note that $a\neq \frac{3}{2}$ for any finite $\lambda$ because $a\to \frac{3}{2}$ as $\lambda \to \infty$.
Since the constant $\kappa$ can be rewritten as $\kappa =\frac{8\pi}{a}$, $a\neq 0$ is required.
When $\lambda =0$, i.e., $a=1$, these equations reduce to those of general relativity.

In general, the metric of a static and spherically symmetric spacetime can be written as
\begin{eqnarray}
ds^2=-e^{2F(r)} dt^2+e^{2G(r)} dr^2 +r^2 (d\theta^2 +\sin^2 \theta d\varphi^2) \label{metric}
\end{eqnarray}
via an appropriate transformation of coordinates,
where $(t,r,\theta,\varphi)$ are the spherical coordinates.
In addition, the metric (\ref{metric}) can be transformed into
\begin{eqnarray}
ds^2=-f(\tr) dt^2+\dfrac{1}{f(\tr)} d\tr^2 +h^2(\tr) (d\theta^2 +\sin^2 \theta d\varphi^2) \label{metric2}
\end{eqnarray}
by a further suitable coordinate transformation.

For the second metric (\ref{metric2}), the following components are obtained:
\begin{eqnarray}
R_{tt}-\dfrac{1}{2} g_{tt} R &=&(h^{-2} -f' h^{-1}h' -f h^{-2}h'^2 -2fh^{-1} h'')f, \\
R_{\tr \tr}-\dfrac{1}{2} g_{\tr \tr} R &=&-(h^{-2} -f' h^{-1}h' -f h^{-2}h'^2)f^{-1}, \\
R_{AB}-\dfrac{1}{2} g_{AB} R &=&\left( \dfrac{1}{2} f'' +f h^{-1} h''+f' h^{-1}h' \right) h^2 \sigma_{AB},
\end{eqnarray}
where the prime denotes a derivative with respect to $\tr$.
Here, the indices $A,B$ denote the angular coordinates, and $\sigma_{AB}$ is the metric of a 2-dimensional unit sphere.

In this paper, we consider a perfect fluid whose stress-energy tensor is given by
\begin{eqnarray}
T_{\mu \nu}=(\rho (\tr) +p(\tr))u_{\mu} u_{\nu} +p(\tr)g_{\mu \nu},
\end{eqnarray}
where $\rho (\tr)$ is the energy density, $p(\tr)$ is the pressure, and $u^{\mu}$ is the four-velocity.
Then, the non-zero components of $T_{\mu \nu}-\frac{a-1}{2} g_{\mu \nu} T$ take the following form:
\begin{eqnarray}
T_{tt} -\dfrac{a-1}{2} g_{tt} T &=& -\dfrac{1}{2} ((a-3)\rho -3(a-1)p) f, \\
T_{\tr \tr} -\dfrac{a-1}{2} g_{\tr \tr} T &=& \dfrac{1}{2} ((a-1)\rho -(3a-5)p) f^{-1}, \\
T_{AB} -\dfrac{a-1}{2} g_{AB} T &=& \dfrac{1}{2} ((a-1)\rho -(3a-5)p) h^2 \sigma_{AB}.
\end{eqnarray}

Here, we replace $\kappa \rho$ and $\kappa p$ as $\rho$ and $p$, respectively,
and adopt the equation of state $p=w\rho$, where $w$ is a parameter.

The gravitational field equation (\ref{Eineq-a}) yields
\begin{eqnarray}
&&h^{-2} -f' h^{-1}h' -f h^{-2}h'^2 -2fh^{-1} h'' = \dfrac{1}{2} ((3w-1)a-3(w-1))\rho, \label{eq1} \\
&&-(h^{-2} -f' h^{-1}h' -f h^{-2}h'^2) = -\dfrac{1}{2} ((3w-1)a-(5w-1))\rho, \label{eq2} \\
&&\dfrac{1}{2} f'' +f h^{-1} h''+f' h^{-1}h'=-\dfrac{1}{2} ((3w-1)a-(5w-1))\rho, \label{eq3}
\end{eqnarray}
while the non-conservation law for the stress-energy tensor (\ref{energy-a}) gives
\begin{eqnarray}
((3w-1)a-(5w-1))\rho' -(w+1)f^{-1} f' \rho =0. \label{eq4}
\end{eqnarray}

The curvature invariants are given by
%Kretschmann invariant
\begin{eqnarray}
R_{\mu \nu \sigma \rho} R^{\mu \nu \sigma \rho} =f''^2+2\left( \dfrac{2h''}{h}f +\dfrac{h'}{h} f' \right)^2
+2\left( \dfrac{h'}{h} \right)^2 f'^2 +4\left( \left( \dfrac{h'}{h} \right)^2 f -\dfrac{1}{h^2} \right)^2, \label{Kret}
\end{eqnarray}
\begin{eqnarray}
R_{\mu \nu} R^{\mu \nu} =\dfrac{1}{4} (f''+2f'h^{-1}h')^2 +\left( \dfrac{1}{2} (f''+2f'h^{-1}h')+2fh^{-1}h'' \right)^{2}
\nonumber \\
+2(h^{-2} -f'h^{-1}h' -f(h^{-1}h''+h^{-2}h'^2))^2,  \label{RR}
\end{eqnarray}
\begin{eqnarray}
R=-2\left( \dfrac{1}{2} f''+2f'h^{-1} h'+f(2h^{-1}h''+h^{-2}h'^2)-h^{-2} \right). \label{R}
\end{eqnarray}

%======================================%
%<<<<<<<<<<<< SECTION III  >>>>>>>>>>>>%
%======================================%

\section{Exact solutions} \label{sec-sol}

When solving the system of differential equations (\ref{eq1})--(\ref{eq4}),
we need to consider a case-by-case analysis during the transformation of the equations depending on the parameters $a$ and $w$,
which allows us to obtain analytical solutions in several cases.
Specifically, for $a\neq \frac{3}{2}$ (see Ref.~\cite{OVF2016} for $a= \frac{3}{2}$), it is necessary to consider the cases $w=\frac{1}{3}$ and $w=-1$,
and when $w\neq \frac{1}{3}, -1$, the case $a=\frac{5w-1}{3w-1}$ must be taken into account.
Furthermore, one can see that the relations $a=\frac{11w-1}{2(3w-1)}$ and $a=\frac{6w}{3w-1}$ also play specific roles
by observing the transformation of the equations.

Among these, the following four cases all yield exact solutions:
(i)~$w=-1$, (ii)~$a=\frac{5w-1}{3w-1}$ with $w\neq \frac{1}{3},-1$, 
(iii)~$a= \frac{11w-1}{2(3w-1)}$ with $w\neq \frac{1}{3}, -1$, and 
(iv)~$a= \frac{6w}{3w-1}$ with $w\neq \frac{1}{3}, -1$.
Therefore, we focus on these four cases in this paper.
In each case, a corresponding term vanishes during the transformation of the equations, 
which in turn enables us to obtain exact solutions.
As will be seen later, these specific relations between $a$ and $w$ play important roles.

We consider Cases (i), (ii), and (iii) in this section, and Case (iv) in the next section.
This is because Case (iv) yields five solutions of two different types.

\subsection{Case $w=-1$}

When $w=-1$, we can derive $\rho'=0$ from Eq. (\ref{eq4}) for any $a\neq \frac{3}{2}$ 
\footnote{
When $w=-1$, the relation $a=\frac{5w-1}{3w-1}$ gives $a=\frac{3}{2}$, hence $a\neq \frac{5w-1}{3w-1}$ holds.
}.
Therefore, we obtain $\rho =\rho_0$, where $\rho_0$ is a constant.

In this case, Eqs. (\ref{eq1}) and (\ref{eq2}) yield
\begin{eqnarray}
2fh^{-1}h''=0,
\end{eqnarray}
which gives
\begin{eqnarray}
h=\rho_1 \tr +\rho_2,
\end{eqnarray}
where $\rho_1, \rho_2$ are constants.
Hence, Eq. (\ref{eq2}) yields
\begin{eqnarray}
f=\dfrac{1}{\rho_1^2} \left( 1-\dfrac{\rho_2 -\rho_1^2 \rho_3}{\rho_1 \tr +\rho_2} 
+\dfrac{(2a-3)\rho_0}{3} (\rho_1 \tr +\rho_2)^2 \right),
\end{eqnarray}
where $\rho_3$ is a constant.
Here, the obtained $f,h$, and $\rho$ also satisfy Eq. (\ref{eq3}).

As a result, $\rho=\rho_0$ and $p=-\rho_0$ hold, and the metric takes the form
\begin{eqnarray}
ds^2 &=&-\dfrac{1}{\rho_1^2} \left( 1-\dfrac{\rho_2 -\rho_1^2 \rho_3}{\rho_1 \tr +\rho_2} 
+\dfrac{(2a-3)\rho_0}{3} (\rho_1 \tr +\rho_2)^2 \right) dt^2 \nonumber \\
&&~~~~+\left[ \dfrac{1}{\rho_1^2} \left( 1-\dfrac{\rho_2 -\rho_1^2 \rho_3}{\rho_1 \tr +\rho_2} 
+\dfrac{(2a-3)\rho_0}{3} (\rho_1 \tr +\rho_2)^2 \right) \right]^{-1} d\tr^2 \nonumber \\
&&~~~~+(\rho_1 \tr +\rho_2)^2 (d\theta^2 +\sin^2 \theta d\varphi^2).
\end{eqnarray}
This form reproduces the Schwarzschild-(anti) de Sitter solution.

To examine the energy conditions, we temporarily restore the constant $\kappa =\frac{8\pi}{a}$ 
into the expressions for $\rho$ and $p$.
By choosing
\begin{eqnarray}
\rho_0 =-\dfrac{\Lambda}{2a-3},
\end{eqnarray}
where $\Lambda$ is the cosmological constant, the metric becomes independent of the Rastall parameter, 
while the energy density and pressure depend on it.
In this case, 
\begin{eqnarray}
8\pi \rho =-\dfrac{a}{2a-3} \Lambda,~~~~8\pi (\rho +p)=0,~~~~8\pi (\rho +3p)=\dfrac{2a}{2a-3} \Lambda
\end{eqnarray}
hold. 
Thus, as in general relativity, the weak and strong energy conditions are not satisfied simultaneously,
and since $w=-1$, the dominant energy condition is always fulfilled whenever the weak energy condition is satisfied.

\subsection{Case $a=\dfrac{5w-1}{3w-1}$ with $w\neq \dfrac{1}{3},-1$}

Let us consider the case $a=\frac{5w-1}{3w-1}$ with $w\neq \frac{1}{3},-1$ 
\footnote{
If $w=\frac{1}{3}$, $a=\frac{5w-1}{3w-1}$ diverges, i.e., $\lambda =\frac{1}{4\kappa}$.
On the other hand, $w=-1$ gives $a=\frac{3}{2}$, which contradicts the condition $a\neq \frac{3}{2}$.
}.
Since $a=1$ in general relativity, this case corresponds to $w=0$.

In this case, Eq. (\ref{eq4}) implies that either $f'=0$ or $\rho =0$.
If $\rho =0$, this corresponds to the vacuum case.
Hence, we are forced to choose $f'=0$, which means that $f$ is a constant.
However, by substituting $f=$ constant into Eqs. (\ref{eq1})--(\ref{eq3}), we can see that $\rho =0$ must hold.
This means that only the trivial solution exists.
In other words, the spacetime is Minkowski, and there is no non-trivial, static, and spherically symmetric solution 
with a perfect fluid.

\subsection{Case $a= \dfrac{11w-1}{2(3w-1)}$ with $w\neq \dfrac{1}{3}, -1$}

In this subsection, let us consider the case $a= \frac{11w-1}{2(3w-1)}$ with $w\neq \frac{1}{3}, -1$.
Since $a=1$ in general relativity, this case corresponds to $w=-\frac{1}{5}$.

Equation (\ref{eq4}) then implies
\begin{eqnarray}
\rho =l_0 f^2, \label{S-rho}
\end{eqnarray}
where $l_0$ is a constant.
Here, we must have $l_0\neq 0$ since $l_0=0$ would correspond to a vacuum spacetime.

From Eqs. (\ref{eq1}) and (\ref{eq2}),
we obtain 
\begin{eqnarray}
2fh''=-(w+1) \rho h. \label{S-fh}
\end{eqnarray}
Therefore, this equation with Eq. (\ref{eq3}) gives us
\begin{eqnarray}
(fh)''=0,
\end{eqnarray}
which yields 
\begin{eqnarray}
f=(l_1\tr+l_2)h^{-1}, \label{S-f}
\end{eqnarray}
where $l_1, l_2$ are constants.

Hence, from Eqs. (\ref{S-rho}), (\ref{S-fh}) and (\ref{S-f}),
we obtain
\begin{eqnarray}
h=-\dfrac{w+1}{12} l_0 l_1\tr^3-\dfrac{w+1}{4} l_0l_2 \tr^2 +l_3 \tr+l_4,
\end{eqnarray}
which yields
\begin{eqnarray}
f=\dfrac{l_1\tr+l_2}{-\frac{w+1}{12} l_0 l_1\tr^3-\frac{w+1}{4} l_0l_2 \tr^2 +l_3 \tr+l_4},
\end{eqnarray}
where $l_3, l_4$ are constants.
Here, $l_0, l_1, l_2,$ and $l_3$ must satisfy the relation
\begin{eqnarray}
1-l_1l_3=\dfrac{1}{4}(w+1) l_0l_2^2.
\end{eqnarray}

Semiz obtained the solution for the case $w=-\frac{1}{5}$ in general relativity \cite{Semiz2022}.
The geometry is analyzed in Ref. \cite{FBF2022}.
Our solution has the same form as Semiz's solution 
\footnote{
Semiz obtained two solutions: the generic case and the particular case.
The particular case corresponds to $l_1=0$ in our solution.
}.

To examine the energy conditions, we temporarily restore the constant $\kappa =\frac{8\pi}{a}$ 
into the expressions for $\rho$ and $p$.
By choosing
\begin{eqnarray}
l_0 &=& \dfrac{4}{w+1} l_5 \nonumber \\
&=& \dfrac{6a-11}{2a-3} l_5,
\end{eqnarray}
where $l_5$ is a constant satisfying $1-l_1 l_3=l_2^2 l_5$,
the metric becomes independent of the Rastall parameter, 
while the energy density and pressure depend on it.
Note that $a\neq 0, \frac{3}{2}, \frac{11}{6}$ since $a= \frac{11w-1}{2(3w-1)}$ and $\kappa =\frac{8\pi}{a}$.
In this case, 
\begin{eqnarray}
8\pi \rho = \dfrac{(6a-11)a}{2a-3} l_5 f^2,~~~~
8\pi (\rho +p)=4a l_5 f^2,~~~~8\pi (\rho +3p)=\dfrac{2a(6a-7)}{2a-3} l_5 f^2
\end{eqnarray}
hold.
The null energy condition requires $a l_5>0$.
Therefore, if $l_5>0$, the weak energy condition is satisfied for $0<a<\frac{3}{2}$ or $a>\frac{11}{6}$,
while the strong energy condition is satisfied for $0<a\leq \frac{7}{6}$ or $a>\frac{3}{2}$, with $a\neq \frac{11}{6}$.
The dominant energy condition is satisfied for $0<a<\frac{3}{2}$ or $a\geq \frac{5}{2}$ 
because both the weak energy condition and the condition $|w|\leq 1$ must be satisfied.
However, if $l_5<0$, all the other energy conditions are also satisfied whenever the null energy condition is satisfied.
Consequently, all the energy conditions are satisfied for $0<a\leq \frac{7}{6}$ or $a\geq \frac{5}{2}$
when $l_5 >0$, and for $a<0$ when $l_5<0$.

%======================================%
%<<<<<<<<<<<< SECTION IV > >>>>>>>>>>>>%
%======================================%

\section{Case $a= \dfrac{6w}{3w-1}$ with $w\neq \dfrac{1}{3}, -1$} \label{sec-exact}

In this section, we consider the case $a= \frac{6w}{3w-1}$ with $w\neq \frac{1}{3}, -1$.
Since $a=1$ in general relativity, this case corresponds to $w=-\frac{1}{3}$.

We find that the exact solutions exhibit richer characteristics than the solutions derived in Sec. \ref{sec-sol},
including wormhole-type geometries with a horizon, standard black hole-type structures, and cyclic closed spacetimes.
Specifically, there are five patterns of exact solutions, depending on the choice of constants.
These patterns can be broadly categorized into two main classes or, in more detail, classified into three types:
the first type possesses a horizon but has no diverging points of curvature invariants;
the second can possess both a horizon and a curvature singularity, depending on the choice of constants;
and the third represents a closed spacetime possessing an infinite number of curvature singularities.
However, the first and second types can be expressed in the same metric form
through an appropriate transformation of coordinates.
This metric exhibits different properties of spacetime depending on the choice of constants,
similar to the metric proposed by Simpson and Visser in Ref. \cite{SV2019}.
Whether a curvature singularity appears or not also depends on the choice of constants.
\\

First, we obtain
\begin{eqnarray}
\rho =k_0 f
\end{eqnarray}
from Eq. (\ref{eq4}), where $k_0$ is a constant.
Therefore, Eqs. (\ref{eq1}) and (\ref{eq2}) yield
\begin{eqnarray}
h''=-\dfrac{w+1}{2}k_0 h. \label{BH-h''}
\end{eqnarray}
Here, $k_0=0$ corresponds to the vacuum case since $\rho =0$.
Hence, we require $k_0\neq 0$.
Furthermore, the condition $(w+1) k_0 \neq 0$ is required since $w\neq -1$.

\subsection{Case $(w+1) k_0<0$}

Let us consider the case $(w+1) k_0<0$.

Equation (\ref{BH-h''}) yields
\begin{eqnarray}
h&=&k_1 e^{\sqrt{-\frac{w+1}{2} k_0} \tr} +k_2 e^{-\sqrt{-\frac{w+1}{2} k_0} \tr} \nonumber \\
&=&(k_1+k_2) \cosh \left( \sqrt{-\frac{w+1}{2} k_0} \tr \right) 
+(k_1-k_2) \sinh \left( \sqrt{-\frac{w+1}{2} k_0} \tr \right), \label{BH-h}
\end{eqnarray}
where $k_1, k_2$ are constants.

\subsubsection{$k_1=0$}

If $k_1=0$, Eq. (\ref{BH-h}) is
\begin{eqnarray}
h=k_2 e^{-\sqrt{-\frac{w+1}{2} k_0} \tr}.
\end{eqnarray}
Then, Eq. (\ref{eq2}) leads to
\begin{eqnarray}
f=-\dfrac{1}{k_2^2} \cdot \dfrac{1}{\left( -(w+1) k_0 \right)} e^{2\sqrt{-\frac{w+1}{2} k_0} \tr} +k_3,
\end{eqnarray}
where $k_3$ is a constant.
The obtained $f,h$, and $\rho$ satisfy Eq. (\ref{eq3}).

There exists a value of $\tr$ such that $f=0$ if $k_3> 0$, 
at which the curvature invariants (\ref{Kret})--(\ref{R}) take finite values.
However, in this case, $f$ diverges and $h \to 0$ as $\tr \to \infty$.

\subsubsection{$k_2=0$}

If $k_2=0$, Eq. (\ref{BH-h}) is
\begin{eqnarray}
h=k_1 e^{\sqrt{-\frac{w+1}{2} k_0} \tr}.
\end{eqnarray}
Then, Eq. (\ref{eq2}) leads to
\begin{eqnarray}
f=-\dfrac{1}{k_1^2} \cdot \dfrac{1}{\left( -(w+1) k_0 \right)} e^{-2\sqrt{-\frac{w+1}{2} k_0} \tr} +k_4,
\end{eqnarray}
where $k_4$ is a constant.
The obtained $f,h$, and $\rho$ satisfy Eq. (\ref{eq3}).

There exists a value of $\tr$ such that $f=0$ if $k_4> 0$, 
at which the curvature invariants (\ref{Kret})--(\ref{R}) take finite values.
In this case, $h$ diverges and $f$ becomes finite as $\tr \to \infty$,
unlike in the $k_1=0$ case.
This means that the exact solution is a black hole solution which has non-divergent curvature invariants.
However, since the condition $(w+1)k_0<0$ is assumed,
this spacetime requires phantom energy with $w<-1$ if $\rho > 0$ is assumed.
Note that, in general relativity, $\rho <0$ is required since $w=-\frac{1}{3}$.

\subsubsection{$k_1 k_2>0$}

Let us consider the case $k_1 k_2>0$ 
\footnote{
Without loss of generality, we can assume $k_1>0$ and $k_2>0$.
}.

Equation (\ref{BH-h}) yields
\begin{eqnarray}
h=2\sqrt{k_1 k_2} \cosh \left( \sqrt{-\frac{w+1}{2} k_0} \tr +\alpha \right),
\end{eqnarray}
where $\alpha$ is a constant satisfying
\begin{eqnarray}
\cosh \alpha =\dfrac{k_1+k_2}{2\sqrt{k_1k_2}}, ~~\sinh \alpha =\dfrac{k_1-k_2}{2\sqrt{k_1k_2}}.
\end{eqnarray}
Therefore, Eq. (\ref{eq2}) leads to
\begin{eqnarray}
f=k_5 \tanh \left( \sqrt{-\frac{w+1}{2} k_0} \tr +\alpha \right) -\dfrac{1}{2k_1k_2 \cdot \left( -(w+1) k_0 \right)},
\end{eqnarray}
where $k_5$ is a constant.
The obtained $f,h$, and $\rho$ satisfy Eq. (\ref{eq3}).

Here, if we assume the asymptotic condition $\displaystyle \lim_{\tr \to \infty} f=1$,
then
\begin{eqnarray}
k_5 =1+\dfrac{1}{2k_1k_2 \cdot \left( -(w+1) k_0 \right)}
\end{eqnarray}
is required.
Therefore, since the spacetime is static and spherically symmetric,
a horizon exists at the value of $\tr$ satisfying $f=0$, i.e.,
\begin{eqnarray}
\tanh \left( \sqrt{-\frac{w+1}{2} k_0} \tr +\alpha \right) 
=\dfrac{1}{1+2k_1k_2 \cdot \left( -(w+1) k_0 \right)}. \label{BH-f=0-1}
\end{eqnarray}
Since $k_1k_2>0$ and $(w+1)k_0<0$ are assumed, the inequality
\begin{eqnarray}
0<\dfrac{1}{1+2k_1k_2 \cdot \left( -(w+1) k_0 \right)}<1
\end{eqnarray}
is satisfied.
Hence, there exists a value of $\tr$ that satisfies Eq. (\ref{BH-f=0-1}).
We can also see that the curvature invariants (\ref{Kret})--(\ref{R}) are finite for all values of $\tr$,
that is, the exact solution is a black hole solution which has non-divergent curvature invariants.
However, since the condition $(w+1)k_0<0$ is assumed,
this spacetime requires phantom energy with $w<-1$ if $\rho > 0$ is assumed.
Note that, in general relativity, $\rho <0$ is required since $w=-\frac{1}{3}$.

\subsubsection{$k_1 k_2<0$}

Next, let us consider the case $k_1 k_2<0$ 
\footnote{
Without loss of generality, we can assume $k_1>0$ and $k_2<0$.
}.

Equation (\ref{BH-h}) yields
\begin{eqnarray}
h=2\sqrt{-k_1 k_2} \sinh \left( \sqrt{-\frac{w+1}{2} k_0} \tr +\beta \right),
\end{eqnarray}
where $\beta$ is a constant satisfying
\begin{eqnarray}
\cosh \beta =\dfrac{k_1-k_2}{2\sqrt{-k_1k_2}}, ~~\sinh \beta =\dfrac{k_1+k_2}{2\sqrt{-k_1k_2}}.
\end{eqnarray}
Therefore, Eq. (\ref{eq2}) leads to
\begin{eqnarray}
f=\dfrac{k_6}{\tanh \left( \sqrt{-\frac{w+1}{2} k_0} \tr +\beta \right)} 
+\dfrac{1}{2(-k_1k_2) \cdot \left( -(w+1) k_0 \right)},
\end{eqnarray}
where $k_6$ is a constant.
The obtained $f,h$, and $\rho$ satisfy Eq. (\ref{eq3}).

Here, if we assume the asymptotic condition $\displaystyle \lim_{\tr \to \infty} f=1$,
then
\begin{eqnarray}
k_6 =1-\dfrac{1}{2(-k_1k_2) \cdot \left( -(w+1) k_0 \right)}
\end{eqnarray}
is required.
Therefore, since the spacetime is static and spherically symmetric,
a horizon exists at the value of $\tr$ satisfying $f=0$, i.e.,
\begin{eqnarray}
\tanh \left( \sqrt{-\frac{w+1}{2} k_0} \tr +\beta \right) 
=1-2(-k_1k_2) \cdot \left( -(w+1) k_0 \right). \label{BH-f=0-2}
\end{eqnarray}

Since $k_1k_2<0$ and $(w+1)k_0<0$ are assumed, the inequality
\begin{eqnarray}
1-2(-k_1k_2) \cdot \left( -(w+1) k_0 \right) <1
\end{eqnarray}
is satisfied.
Hence, for an appropriate choice of the constants, there exists a value of $\tr$ that satisfies Eq. (\ref{BH-f=0-2}).
We can also see that the curvature invariants (\ref{Kret})--(\ref{R}) diverge at
\begin{eqnarray}
\tr=\dfrac{-\beta}{\sqrt{-\frac{w+1}{2} k_0}}.
\end{eqnarray}
At this value, $h$ vanishes and $f$ diverges.

\subsection{Case $(w+1) k_0>0$}

Finally, let us consider the case $(w+1) k_0>0$.

Equation (\ref{BH-h''}) yields
\begin{eqnarray}
h&=&k_7 \cos \left(\sqrt{\frac{w+1}{2} k_0} \tr \right) +k_8 \sin \left(\sqrt{\frac{w+1}{2} k_0} \tr \right) \nonumber \\
&=&\sqrt{k_7^2+k_8^2} \sin \left( \sqrt{\frac{w+1}{2} k_0} \tr +\gamma \right), \label{BH-y-2}
\end{eqnarray}
where $k_7, k_8$ are constants, and $\gamma$ is a constant satisfying
\begin{eqnarray}
\cos \gamma =\dfrac{k_8}{\sqrt{k_7^2+k_8^2}}, ~~\sin \gamma =\dfrac{k_7}{\sqrt{k_7^2+k_8^2}}.
\end{eqnarray}
Therefore, Eq. (\ref{eq2}) leads to
\begin{eqnarray}
f=\dfrac{k_9}{\tan \left( \sqrt{\frac{w+1}{2} k_0} \tr +\gamma \right)} +\dfrac{1}{(k_7^2+k_8^2) \cdot \frac{w+1}{2} k_0},
\end{eqnarray}
where $k_9$ is a constant.
The obtained $f,h$, and $\rho$ satisfy Eq. (\ref{eq3}).

We can see that the curvature invariants (\ref{Kret})--(\ref{R}) diverge at the values of $\tr$ that satisfy
\begin{eqnarray}
\sqrt{\frac{w+1}{2} k_0} \tr +\gamma =n \pi,
\end{eqnarray}
where $n$ is an integer.
This means that there are an infinite number of curvature singularities.

\subsection{Coordinate transformation}

The five exact solutions obtained in the previous two subsections are expressed as the following metrics, respectively:
\begin{eqnarray}
(\mathrm{I})~&&ds^2=
-\left[ k_3 -\dfrac{1}{k_2^2 (-(w+1)k_0)} e^{2\sqrt{-\frac{w+1}{2} k_0}\tr} \right] dt^2 \nonumber \\
&&~~~~~~~~~~~~
+\left[ k_3 -\dfrac{1}{k_2^2 (-(w+1)k_0)} e^{2\sqrt{-\frac{w+1}{2} k_0}\tr} \right]^{-1} d\tr^2 \nonumber \\
&&~~~~~~~~~~~~
+k_2^2 e^{-2\sqrt{-\frac{w+1}{2} k_0}\tr} (d\theta^2 +\sin^2 \theta d\varphi^2),
\end{eqnarray}
\begin{eqnarray}
(\mathrm{II})~&&ds^2=
-\left[ k_4 -\dfrac{1}{k_1^2 (-(w+1)k_0)} e^{-2\sqrt{-\frac{w+1}{2} k_0}\tr} \right] dt^2 \nonumber \\
&&~~~~~~~~~~~~
+\left[ k_4 -\dfrac{1}{k_1^2 (-(w+1)k_0)} e^{-2\sqrt{-\frac{w+1}{2} k_0}\tr} \right]^{-1} d\tr^2 \nonumber \\
&&~~~~~~~~~~~~
+k_1^2 e^{2\sqrt{-\frac{w+1}{2} k_0}\tr} (d\theta^2 +\sin^2 \theta d\varphi^2),
\end{eqnarray}
\begin{eqnarray}
(\mathrm{III})~&&ds^2=
-\left[ k_5 \tanh \left( \sqrt{-\frac{w+1}{2} k_0} \tr +\alpha \right) 
-\dfrac{1}{2k_1k_2 \cdot \left( -(w+1) k_0 \right)} \right] dt^2 \nonumber \\
&&~~~~~~~~~~~~
+\left[ k_5 \tanh \left( \sqrt{-\frac{w+1}{2} k_0} \tr +\alpha \right) 
-\dfrac{1}{2k_1k_2 \cdot \left( -(w+1) k_0 \right)} \right]^{-1} d\tr^2 \nonumber \\
&&~~~~~~~~~~~~
+4k_1 k_2 \cosh^2 \left( \sqrt{-\frac{w+1}{2} k_0} \tr +\alpha \right) (d\theta^2 +\sin^2 \theta d\varphi^2),
\end{eqnarray}
\begin{eqnarray}
(\mathrm{IV})~&&ds^2=
-\left[ \dfrac{k_6}{\tanh \left( \sqrt{-\frac{w+1}{2} k_0} \tr +\beta \right)} 
+\dfrac{1}{2(-k_1k_2) \cdot \left( -(w+1) k_0 \right)} \right] dt^2 \nonumber \\
&&~~~~~~~~~~~~
+\left[ \dfrac{k_6}{\tanh \left( \sqrt{-\frac{w+1}{2} k_0} \tr +\beta \right)} 
+\dfrac{1}{2(-k_1k_2) \cdot \left( -(w+1) k_0 \right)} \right]^{-1} d\tr^2 \nonumber \\
&&~~~~~~~~~~~~
+4(-k_1 k_2) \sinh^2 \left( \sqrt{-\frac{w+1}{2} k_0} \tr +\beta \right) (d\theta^2 +\sin^2 \theta d\varphi^2),
\end{eqnarray}
\begin{eqnarray}
(\mathrm{V})~&&ds^2=
-\left[ \dfrac{k_9}{\tan \left( \sqrt{\frac{w+1}{2} k_0} \tr +\gamma \right)} 
+\dfrac{1}{(k_7^2+k_8^2) \cdot \frac{w+1}{2} k_0} \right] dt^2 \nonumber \\
&&~~~~~~~~~~~~
+\left[ \dfrac{k_9}{\tan \left( \sqrt{\frac{w+1}{2} k_0} \tr +\gamma \right)} 
+\dfrac{1}{(k_7^2+k_8^2) \cdot \frac{w+1}{2} k_0} \right]^{-1} d\tr^2 \nonumber \\
&&~~~~~~~~~~~~
+(k_7^2+k_8^2) \sin^2 \left( \sqrt{\frac{w+1}{2} k_0} \tr +\gamma \right) (d\theta^2 +\sin^2 \theta d\varphi^2).
\end{eqnarray}
To make these comparisons, we apply appropriate coordinate transformations to each metric.

The metrics (I) and (II) can be written as 
\begin{eqnarray}
ds^2 &=&-\left( \left( -\dfrac{w+1}{2} k_0 \right) k_3 -\dfrac{1}{2R^2} \right) dT^2
+\dfrac{1}{R^2} 
\left( \left( -\dfrac{w+1}{2} k_0 \right) k_3 -\dfrac{1}{2R^2} \right)^{-1} dR^2 \nonumber \\
&&~~~~
+R^2 (d\theta^2 +\sin^2 \theta d\varphi^2) \label{w=-1/3(1)}
\end{eqnarray}
via appropriate coordinate transformations, respectively.
This metric has a horizon and a curvature singularity, and has the same form as the solution $S_1$ given by Chernin et al. \cite{CSS2002}.

The metrics (III) and (IV) are written as 
\begin{eqnarray}
ds^2 &=&-\left( -\dfrac{w+1}{2} k_0 k_5 \dfrac{\sqrt{R^2-4k_1k_2}}{R} -\dfrac{1}{4k_1k_2} \right) dT^2 \nonumber \\
&&~~~~
+\dfrac{1}{R^2-4k_1k_2} 
\left( -\dfrac{w+1}{2} k_0 k_5 \dfrac{\sqrt{R^2-4k_1k_2}}{R} -\dfrac{1}{4k_1k_2} \right)^{-1} dR^2 \nonumber \\
&&~~~~
+R^2 (d\theta^2 +\sin^2 \theta d\varphi^2) \label{w=-1/3(2)}
\end{eqnarray}
via appropriate coordinate transformations, respectively.
This metric may have a horizon or a curvature singularity depending on the parameters,
and has the same form as the solution $S_{\pm}$ given by Chernin et al. \cite{CSS2002}.

The metric (V) is written as
\begin{eqnarray}
ds^2 &=&-\left( \dfrac{w+1}{2} k_0k_9 \dfrac{\sqrt{(k_7^2+k_8^2)-R^2}}{R} +\dfrac{1}{k_7^2+k_8^2} \right) dT^2 \nonumber \\
&&~~~~
+\dfrac{1}{(k_7^2+k_8^2)-R^2}
\left( \dfrac{w+1}{2} k_0k_9 \dfrac{\sqrt{(k_7^2+k_8^2)-R^2}}{R} +\dfrac{1}{k_7^2+k_8^2} \right)^{-1} dR^2 \nonumber \\
&&~~~~
+R^2 (d\theta^2 +\sin^2 \theta d\varphi^2) \label{w=-1/3(3)}
\end{eqnarray}
via an appropriate coordinate transformation.
This metric has an upper bound on the range of $R$.
In Ref.~\cite{CSS2002}, Chernin et al. commented on this type of metric under case (B) of the solution $S_{\pm}$ 
\footnote{
In Ref.~\cite{CSS2002}, Chernin et al. derived two solutions named $S_2$ and $S_3$ in addition to $S_{\pm}$ and $S_1$.
However, the solutions $S_2$ and $S_3$ are not valid solutions when $w=-\frac{1}{3}$.
Therefore, such solutions do not appear in the present study.
}.

Both metrics (\ref{w=-1/3(2)}) and (\ref{w=-1/3(3)}) are referred to as the Whittaker solution \cite{Maeda2025} 
(see also Ref. \cite{CK2017}).

To examine the energy conditions, we temporarily restore the constant $\kappa =\frac{8\pi}{a}$ 
into the expressions for $\rho$ and $p$.
By choosing $k_3, k_5$, and $k_9$ as
\begin{eqnarray}
\dfrac{2}{w+1} \cdot \dfrac{1}{k_0} =\dfrac{3(a-2)}{2a-3} \cdot \dfrac{1}{k_0}
\end{eqnarray}
for the metrics (\ref{w=-1/3(1)}), (\ref{w=-1/3(2)}), and (\ref{w=-1/3(3)}), respectively,
each corresponding metric becomes independent of the Rastall parameter, 
while the energy density and pressure depend on it.
Note that $a\neq 0, \frac{3}{2}, 2$ since $a= \frac{6w}{3w-1}$ and $\kappa =\frac{8\pi}{a}$.
In this case, 
\begin{eqnarray}
8\pi \rho = a k_0 f,~~~~
8\pi (\rho +p)=\dfrac{2a(2a-3)}{3(a-2)} k_0 f,~~~~8\pi (\rho +3p)=\dfrac{2a(a-1)}{a-2} k_0 f
\end{eqnarray}
hold.
If $k_0>0$, the null energy condition requires $0<a< \frac{3}{2}$ or $a>2$.
Therefore, the weak energy condition is satisfied for the same range of $a$,
while the strong energy condition is satisfied for $0<a \leq 1$ or $a>2$.
The dominant energy condition is satisfied for $0<a< \frac{3}{2}$ or $a \geq 3$ 
because both the weak energy condition and the condition $|w|\leq 1$ must be satisfied.
If $k_0<0$, the null energy condition requires $a<0$ or $\frac{3}{2} <a<2$.
Therefore, the weak and dominant energy conditions are satisfied for $a<0$, 
while the strong energy condition is satisfied for $a<0$ or $\frac{3}{2} < a <2$.
Consequently, all the energy conditions are satisfied for $0<a \leq 1$ or $a\geq 3$ when $k_0>0$, 
and for $a<0$ when $k_0<0$.

%======================================%
%<<<<<<<<<<<< SECTION V  >>>>>>>>>>>>>>%
%======================================%

\section{Summary and Discussion} \label{sec-summary}

In this paper, we assumed the equation of state $p(r)=w\rho (r)$ and 
derived exact solutions for a static and spherically symmetric spacetime with a perfect fluid 
within the framework of Rastall theory.
By determining the relations between the Rastall parameter and the constant $w$, 
we obtained exact solutions that reduce to the counterpart solutions derivable in general relativity
for the values $w=0,-\frac{1}{5}, -\frac{1}{3}$, and $-1$.
We found that this enables us to establish a correspondence with the Rastall parameter for these values of $w$.
In general relativity, it is known that under the same conditions considered in this paper, 
exact Liouvillian solutions exist only for the cases where $w=0,-\frac{1}{6},-\frac{1}{5}, -\frac{1}{3}$, 
and $-1$ \cite{IL2014}.
In this study, however, we successfully obtained the corresponding exact solutions 
within Rastall theory for all these cases except $w=-\frac{1}{6}$.
Moreover, these solutions can be derived simply by identifying the values that cause the coefficients in the differential 
equation to vanish--
that is, by treating the specific cases where constant coefficients 
require separate analysis during the reduction of the Rastall field equations--
without employing any other special methods or assumptions.
Furthermore, regarding the obtained solutions, if we consider only the metric, 
the influence of the Rastall parameter can be eliminated through appropriate choices of arbitrary constants for each case.
Considering our results from a mathematical analysis perspective based on the theorem proved in Ref. \cite{IL2014},
these results seem to support the claim that Rastall theory is equivalent to general relativity.
However, we cannot conclude that the two theories are equivalent based on this analysis alone.
This is because the fact that, for any value of $w$ other than $w=\frac{1}{3}$ and $-1$, 
an appropriate choice of the Rastall parameter yields exact Liouvillian solutions
corresponding to the $w=0,-\frac{1}{5},$ and $-\frac{1}{3}$ cases in general relativity
suggests that the Rastall parameter effectively ``tunes'' the matter fields.
Furthermore, for all the metrics obtained in this paper, 
the fact that the influence of the Rastall parameter appears in $\rho$ and $p$ 
while the metrics reduce to those in general relativity
supports this interpretation.
Therefore, we can consider this to be a candidate for the physical interpretation of the Rastall parameter.
It should be noted that this does not result from the consideration of observational data, 
but rather presents a mathematical possibility for the physical interpretation of the Rastall parameter as matter.

Here, one might notice a similarity between our current results and those of Hansraj et al. \cite{HBC2018}, 
who utilized the Tolman solutions.
In fact, in Ref. \cite{HBC2018}, the metric is the same as that in general relativity, 
while the influence of the Rastall parameter appears in $\rho$ and $p$.
The major difference between the two lies in the derivation process of the obtained solutions.
The Tolman metrics are solutions for a static and spherically symmetric spacetime with a perfect fluid
derived under specific additional assumptions (i.e., ansatzes) on the metric, rather than an equation of 
state \cite{Tolman1939}.
While their solutions are derived using the relation $T_r^r=T_A^A$ resulting from the perfect fluid assumption,
owing to these ansatzes on the metric,
the solutions (i.e., metrics) can be derived simply by solving the differential equation obtained from this relation.
Note that the relation $T_r^r=T_A^A$ is equivalent to $G_r^r=G_A^A$, where $G_{\mu \nu}$ is the Einstein tensor,
in both general relativity and Rastall theory.
Thus, the influence of the Rastall parameter does not appear in their metrics.
On the other hand, in our current study, no additional assumptions are imposed on the metric.
Therefore, the differential equation obtained from the relation $T_r^r=T_A^A$ (i.e., $G_r^r=G_A^A$) cannot be solved in isolation,
and the equation from the $tt-$component of the field equations is also required.
As a result, the influence of the Rastall parameter might also appear in the metric.
For this reason, this implies that our current results carry a fundamentally different significance from those of Hansraj et al.,
even though the metrics we derived can be expressed in forms without the Rastall parameter
by appropriately choosing the arbitrary constants.

Furthermore, in this paper, 
since we have derived exact solutions using the general metric for a static and spherically symmetric spacetime 
without imposing any additional constraints on the metric,
our results are closely related to the uniqueness of the solutions under the respective conditions of the specific relations 
between the Rastall parameter and the constant $w$ of the equation of state $p(r)=w\rho (r)$.
It is thus shown that, under these conditions, the solutions are identical to those in general relativity.
Therefore, further analysis of the spacetimes described by the solutions obtained in this paper 
does not appear to be necessary for resolving the equivalence problem between Rastall theory and general relativity.
Note that the reason a solution corresponding to $w=-\frac{1}{6}$ could not be derived is thought to be that, 
even in general relativity, the derivation method for the $w=-\frac{1}{6}$ case differs from those for other cases, 
involving a distinct theoretical background based on more advanced mathematics, 
such as the Galois extension of differential fields \cite{IL2014, Semiz2022-2}.
Therefore, this equivalence cannot be denied simply on the grounds 
that the exact Liouvillian solution corresponding to $w=-\frac{1}{6}$ was not derived using the present method.
However, in Rastall theory, attempting to derive the solution corresponding to $w=-\frac{1}{6}$ 
or to determine the necessary and sufficient conditions on the value of $w$ for which solutions can be derived
would be an interesting subject for exploring to what extent the assumption of the equation of state $p(r)=w\rho (r)$ 
and the restriction to exact Liouvillian solutions affect the Rastall parameter.

In light of the above considerations, in future work, we aim to investigate the relationship with Rastall theory 
regarding exact Liouvillian solutions corresponding to $w=-\frac{1}{6}$ as well as other values of $w$.
Furthermore, we plan to extend our analysis to cases with different assumptions, 
such as non-spherically symmetric spacetimes, anisotropic fluids, or equations of state beyond $p(r)=w\rho (r)$.
Extending the analysis to $n$-dimensional spacetimes, as investigated in general relativity in Ref.~\cite{MM2024},
would also be an interesting and worthwhile direction.

%
%======================================%
%<<<<<<<<<<<< acknowledgments  >>>>>>>>%
%======================================%
%
\begin{acknowledgments}
Y. T. is grateful to S. Tomizawa, R. Suzuki, and H. Yoshino for helpful suggestions.
The authors thanks H. Maeda for his helpful comments.
The authors used Gemini 3.5 Flash developed by Google to improve the English spelling, grammar, wording, and phrasing in this manuscript.
The authors have carefully reviewed and take full responsibility for all the content.
\end{acknowledgments}

\begin{appendix}
\section{General case of $a$ and $w$} \label{sec-app}

For cases other than those discussed in the main text,
if we set
\begin{eqnarray}
e^{2G(r)}=\left(1-\dfrac{b(r)}{r} \right)^{-1}
\end{eqnarray}
in metric (\ref{metric}), the problem reduces to solving 
a second-order nonlinear ordinary differential equation for $F(r)$.

In this case, the non-zero components of the left-hand side and right-hand side of the field equation (\ref{Eineq-a}) are calculated as
\begin{eqnarray}
R_{tt}-\dfrac{1}{2} g_{tt} R &=&\dfrac{b'}{r^2} e^{2F}, \\
R_{rr}-\dfrac{1}{2} g_{rr} R &=&
\left( 1-\dfrac{b}{r} \right)^{-1} \left[ \dfrac{2}{r} \left( 1-\dfrac{b}{r} \right) F' -\dfrac{b}{r^3} \right], \\
R_{AB}-\dfrac{1}{2} g_{AB} R &=&
r^2 \left[ \left( F''+F'^2 +\dfrac{1}{r}F' \right) \left( 1-\dfrac{b}{r} \right) -\dfrac{b'r-b}{2r^3} (1+rF') \right] \sigma_{AB},
\end{eqnarray}
and
\begin{eqnarray}
T_{tt} -\dfrac{a-1}{2} g_{tt} T &=& -\dfrac{1}{2} ((a-3)\rho -3(a-1)p) e^{2F}, \\
T_{rr} -\dfrac{a-1}{2} g_{rr} T &=& \dfrac{1}{2} ((a-1)\rho -(3a-5)p) \left( 1-\dfrac{b}{r} \right)^{-1}, \\
T_{AB} -\dfrac{a-1}{2} g_{AB} T &=& \dfrac{1}{2} ((a-1)\rho -(3a-5)p) r^2 \sigma_{AB},
\end{eqnarray}
where the prime denotes a derivative with respect to $r$.

Therefore, we obtain the following equations:
\begin{eqnarray}
&&\dfrac{b'}{r^2} = \dfrac{1}{2} ((3w-1)a-3(w-1))\rho, \label{app-eq1} \\
&&\dfrac{2}{r} \left( 1-\dfrac{b}{r} \right) F' -\dfrac{b}{r^3} = -\dfrac{1}{2} ((3w-1)a-(5w-1))\rho, \label{app-eq2} \\
&&\left( F''+F'^2 +\dfrac{1}{r}F' \right) \left( 1-\dfrac{b}{r} \right) -\dfrac{b'r-b}{2r^3} (1+rF') 
=-\dfrac{1}{2} ((3w-1)a-(5w-1))\rho. \label{app-eq3}
\end{eqnarray}
We also obtain 
\begin{eqnarray}
\dfrac{1}{2} ((3w-1)a-(5w-1))\rho' -(w+1)F' \rho =0 \label{app-eq4}
\end{eqnarray}
from the non-conservation law (\ref{energy-a}).

\subsection{Case $w=\dfrac{1}{3}$} \label{app-1}

Equation (\ref{app-eq4}) yields
\begin{eqnarray}
\rho =a_0 e^{-4F},
\end{eqnarray}
where $a_0$ is a constant.
Therefore, Eq. (\ref{app-eq2}) leads to
\begin{eqnarray}
b=\dfrac{r^3}{1+2rF'} \left( \dfrac{2}{r} F'-\dfrac{1}{3}\rho \right).
\end{eqnarray}
From the above two equations along with Eqs. (\ref{app-eq1}) and (\ref{app-eq3}),
we find that $F(r)$ satisfies
\begin{eqnarray}
F''+2F'^2+\dfrac{2}{r} F' +\dfrac{a_0}{3} r^2 \left( F''-2F'^2-\dfrac{6}{r} F' -\dfrac{3}{r^2} \right) e^{-4F}=0.
\end{eqnarray}

\subsection{Case $a\neq \dfrac{5w-1}{3w-1}$ with $w\neq \dfrac{1}{3}, -1$} \label{app-2}

We define $c_1 =\frac{1}{2} ((3w-1)a-(5w-1))$ and $c_2 =\frac{1}{2}((3w-1)a-3(w-1))$.

Equation (\ref{app-eq4}) yields
\begin{eqnarray}
\rho =a_1 e^{\frac{w+1}{c_1} F},
\end{eqnarray}
where $a_1$ is a constant.
Therefore, Eq. (\ref{app-eq2}) leads to
\begin{eqnarray}
b=\dfrac{r^3}{1+2rF'} \left( \dfrac{2}{r} F' +c_1 \rho \right).
\end{eqnarray}
From the above two equations along with Eqs. (\ref{app-eq1}) and (\ref{app-eq3}),
we find that $F(r)$ satisfies
\begin{eqnarray}
&&F''+2F'^2+\dfrac{2}{r} F' -a_1 c_1 r^2 \left( F''+F'^2-\dfrac{3}{2r} F' -\dfrac{3}{2r^2} \right) e^{\frac{w+1}{c_1} F} 
\nonumber \\
&&~~~~-a_1 c_2 r^2 \left( F'^2+\dfrac{3}{2r} F' +\dfrac{1}{2r^2} \right) e^{\frac{w+1}{c_1} F} =0. \label{eq-app}
\end{eqnarray}

\end{appendix}

%---------   References   ---------%

\end{document}